\documentclass[sigconf,nonacm=true]{acmart}

\usepackage{multirow} 
\usepackage[most]{tcolorbox}
\usepackage{soul}
\usepackage{cleveref}
\usepackage{enumitem}
\usepackage{bbding}

\definecolor{darkred}{RGB}{180,30,30}
\definecolor{softpurple}{RGB}{120,90,160}
\definecolor{darkpurple}{RGB}{65,45,105}
\newcommand{\ba}[1]{\textbf{#1}}
\newcommand{\bs}[1]{#1\rlap{$^{*}$}}
\newcommand{\itt}[1]{\textit{#1}}
\newcommand{\bi}[1]{\textbf{\textit{#1}}}
\newcommand{\bu}[1]{\underline{#1}}

\newcommand{\greenCheck}{\textcolor{green!60!black}{\CheckmarkBold}}
\newcommand{\redCross}{\textcolor{red!80!black}{\XSolidBrush}}

\AtBeginDocument{%
  \providecommand\BibTeX{{%
    \normalfont B\kern-0.5em{\scshape i\kern-0.25em b}\kern-0.8em\TeX}}}

\acmISBN{}

\graphicspath{{./images/}}

\begin{document}

\title{Keep it Simple! A Multiple-Choice Task can Benefit
Generative Recommendation
}

\author{Wei-Hsiang Huang}
\orcid{0009-0009-0901-9668}
\affiliation{%
  \institution{National Taiwan University}
  \city{Taipei}
  \country{Taiwan}
}
\email{ddavid200133@gmail.com}

\author{Jyun-Yu Jiang}
\affiliation{%
  \institution{Coupang}
  \city{California}
  \country{USA}
}
\email{jyunyu.jiang@gmail.com}

\author{Pu-Jen Cheng}
\affiliation{%
  \institution{National Taiwan University}
  \city{Taipei}
  \country{Taiwan}
}
\email{pjcheng@csie.ntu.edu.tw}



\begin{abstract}

Sequential recommendation has recently witnessed growing interest in generative retrieval.
One of the most popular approaches is to autoregressively generate the semantic identifiers (SIDs) of items as recommendations.
Existing efforts mainly 
rely on semantic alignment,
sophisticated architectural designs,
and extensive hyperparameter tuning.
However, the conventional learning framework for generative models could have foundational limitations.
Specifically, most generative models directly predict target items without explicitly learning negative compositions and relative ranking signals.
This is not well aligned with the philosophy of Bayesian personalized ranking~(BPR), which emphasizes the importance of relative rankings between positive and negative samples.

In this paper, we show that adding a simple multiple-choice auxiliary task into the learning objective can effectively address this issue.
With minimal adaptation, we introduce \textit{Task Identifiers} and \textit{Candidate Constructions} to synthesize training data for the multiple-choice auxiliary task.
In addition, we provide theoretical analysis of the fundamental benefits of the multiple-choice auxiliary task from the perspectives of information theory and compositional generalization.
Extensive experiments on four publicly available benchmark datasets show that our approach can improve TIGER, a conventional generative recommendation model, by 4.9\% to 79\% in NDCG@10.

\end{abstract}




\maketitle

\section{Introduction}

Sequential recommendation~\cite{kang2018selfattentivesequentialrecommendation,sun2019bert4recsequentialrecommendationbidirectional,10.1145/3340531.3411954,geng2022recommendation,10.1145/3583780.3615017,cao2024aligninglargelanguagemodels,wang2024rdrec,rajput2023recommendersystemsgenerativeretrieval,10.1145/3627673.3679569} has long been a critical
domain within recommendation systems.
In general, these models recommend the next item by capturing the underlying intents within temporal sequences of user interactions.
Due to the sequential nature,
several recent studies~\cite{geng2022recommendation,10.1145/3583780.3615017,cao2024aligninglargelanguagemodels,wang2024rdrec,rajput2023recommendersystemsgenerativeretrieval,10.1145/3627673.3679569}
focus on the generative paradigm. 
Within this space, a significant focus~\cite{rajput2023recommendersystemsgenerativeretrieval,zheng2024adaptinglargelanguagemodels,10.1145/3627673.3679569,10.1145/3774904.3792802,10.1145/3726302.3729989} 
has emerged on 
the use of Semantic Identifiers (SIDs).
This usually includes two steps of development,
\textit{SID Generation} and \textit{Generative 
Retrieval}. First, SIDs are generated by encoding 
item semantics with hierarchical coding, such as RQ-VAE~\cite{zeghidour2021soundstream,rajput2023recommendersystemsgenerativeretrieval} or RK-Means~\cite{10.1145/3746252.3761502,deng2025onerec,10.1145/3746252.3761612}. 
Subsequently, the chronological items interaction is 
mapped into a sequence of item SIDs and fed into the generative recommender. The next item SID
is then generated via the autoregressive process.

Based on this architecture,
existing efforts 
focus on integrating and aligning
collaborative information into language 
models~\cite{zheng2024adaptinglargelanguagemodels,10.1145/3774904.3792802,li2024semanticconvergenceharmonizingrecommender}, or
maintaining more effective SID 
representations~\cite{10.1145/3627673.3679569,10.1145/3726302.3729989,10.1145/3774904.3792802} 
to better align with recommenders.
Despite their effectiveness,
these models often rely on multiple alignment tasks, complex 
architectural designs, or 
extensive hyperparameter tuning.
These issues could stem from the original design of the learning framework for generative models.
First, the retrieval task involves substantial ambiguity when models are learned to generate the next item directly from the entire item corpus.
In other words, models during training would only be asked to memorize fixed patterns.
In particular, these models could suffer from out-of-distribution patterns, such as unseen compositions,
as these patterns are rarely observed during training.
Moreover, recent 
studies~\cite{10.1145/3627673.3679569,10.1145/3773771}
point out
that the 
collaborative signals
are beneficial to be added into the constructions of SIDs or 
model architectures, 
thereby bridging the gap between semantics and 
behavioral interaction patterns. 
However, it is arguable that the collaborative information
can be naturally captured
when training generative retrieval models on sequential data.
From this aspect, retrieval
models are expected to retain sufficient collaborative 
knowledge, without the need
of introducing external models
and extensive hyperparameter tuning.

These observations motivate us to revisit the design 
of the sequential training objective itself. 
In particular, we note that the current approach
only defines the objective to
align sequences with 
ground-truth target items, 
while leaving negative compositions unspecified.
This idea is not well-aligned with Bayesian personalized ranking~(BPR)~\cite{rendle2012bprbayesianpersonalizedranking}, which emphasizes the importance of
relative rankings between positive and
negative samples.
To mitigate these challenges, we propose to add a simple 
multiple-choice
auxiliary task to the learning objective as a natural and effective remedy.
Specifically, the multiple-choice auxiliary task can explicitly present a subset of candidate items to encourage models to learn both positive and negative signals.
This additional task can be seamlessly
fit into the original sequential models,
such as TIGER~\citep{rajput2023recommendersystemsgenerativeretrieval},
even without relying on pretrained language weights. 
The whole framework remains simple and elegant without introducing any complicated training strategy.

In this paper, we propose to integrate the multiple-choice task into generative sequential recommendation.
Extensive experiments validate the effectiveness and investigate the underlying principles about how this auxiliary task addresses the aforementioned limitations.
Our contributions can be summarized as follows:
\begin{itemize}[leftmargin=*]
    \item We provide a simple manner to integrate the multiple-choice auxiliary task into generative recommendation models, along
    with theoretical analysis of its advantages.
    \item We show that the
    multiple-choice auxiliary task can successfully 
    mitigate the limitations and achieve better recommendation performance.
    \item We conduct a series of in-depth analysis, including the exploration of the underlying factors behind the performance, the relative importance of the main sequential task and
    the auxiliary task, and the impacts of different candidate sampling strategies.
\end{itemize}

\section{Related Work}

\subsection{Generative Recommendation System}
Generative recommendation has been
a widely studied area~\cite{geng2022recommendation,10.1145/3583780.3615017,rajput2023recommendersystemsgenerativeretrieval,10.1145/3624918.3625339,zheng2024adaptinglargelanguagemodels,10.1145/3627673.3679569,10.1145/3726302.3729989,wang2024rdrec}.
In recent years, 
research has increasingly 
focused on representing 
items using semantic identifiers (SIDs),
which are constructed through vector 
quantization (e.g., RQ-VAE), as
first proposed by TIGER~\cite{rajput2023recommendersystemsgenerativeretrieval}.
Following this line of work, LC-Rec~\cite{zheng2024adaptinglargelanguagemodels}
integrates the ideas of SID on LLMs, and 
designs tasks to bridge the gap between
language and recommendation information.
LETTER~\cite{10.1145/3627673.3679569} formulates
better SID representations by considering
collaborative signals 
and controlling the
code embeddings diversity. ETEGRec~\cite{10.1145/3726302.3729989}
learns SIDs
in an end-to-end manner, 
allowing the item tokenizer and recommender to
be jointly optimized through the alignment
between sequences and items, as well
as preferences and semantics.

These approaches have been shown to improve
the generative recommendation system.
However, they often rely on 
multiple alignment objectives 
between language, collaborative signals, 
and SID representations, which introduce additional hyperparameters 
and therefore increase optimization complexity.
Instead, our work directly utilizes the
natural property of sequence-to-sequence model
architectures to bridge the gaps
between pretrained SIDs and collaborative
knowledge.

\subsection{Auxiliary Tasks on Sequential Recommendation}
Recently, auxiliary tasks have been widely 
adopted in sequential 
recommendation systems to enhance 
model performance and representation learning.
$S^3$-Rec~\cite{10.1145/3340531.3411954} 
and UniSRec~\cite{10.1145/3534678.3539381}
leverage auxiliary tasks, 
such as cloze, attributes matching, 
and sequence-sequence contrastive learning, for model 
pretraining.
Several LLM-based recommenders
focus on multi-task learning (MTL)~\cite{geng2022recommendation, 10.1145/3583780.3615017},
or 
design auxiliary tasks
to align recommendation knowledge with
language semantics~\cite{zheng2024adaptinglargelanguagemodels, wang2024rdrec, cao2024aligninglargelanguagemodels,li2024semanticconvergenceharmonizingrecommender}.  
Among them, P5~\cite{geng2022recommendation} expresses items
as their naive IDs and includes the multiple-choice task 
as one of the primary objectives.
\cite{cao2024aligninglargelanguagemodels} extends P5 by
introducing additional auxiliary tasks, such as the BPR task,
and represents items using both naive IDs and titles.
On the other hand, \cite{li2024semanticconvergenceharmonizingrecommender} focuses on
SID-based item representations and constructs the negative sampling task to learn from items that users are unlikely to interact with.

Despite their effectiveness,
prior studies either rely on language-based item identifiers
or introduce negative signals only through the training labels
without explicit comparison between positive and negative items.
In our work, we identify
the lack of comparative signals between
items in recommendation sequences
and investigate how the multiple-choice task can
benefit the SID-based generative recommendation system.
Notably, unlike previous works, our proposed framework does not rely on pretrained language weights during generative process.
We compare our work with several related studies that
contain the idea of multiple-choice task or negative sampling
in Table~\ref{tab:related-study-comparison}.

\begin{table}
\centering
\caption{
Comparison of our work with related studies that contain the multiple-choice task or negative sampling, where 
``Explicit Comparison'' means the model
explicitly contrasts positive items against negative ones, and
``LLM free'' indicates the proposed method does not rely on pretrained language weights during generative process.
}
\label{tab:related-study-comparison}
\resizebox{\linewidth}{!}{
\begin{tabular}{lcccc}
\toprule
Method & SID-based & Explicit Comparison  & LLM Free \\
\midrule
P5~\cite{geng2022recommendation} & \redCross & \greenCheck & \redCross \\
Cao et al.~\cite{cao2024aligninglargelanguagemodels} & \redCross & \greenCheck & \redCross \\
Li et al.~\cite{li2024semanticconvergenceharmonizingrecommender} & \greenCheck & \redCross & \redCross \\
\midrule
Ours & \greenCheck & \greenCheck &  \greenCheck \\
\bottomrule
\end{tabular}}
\end{table}

\subsection{Compositional Generalization}
In various domains, such as Natural Language Processing (NLP) and 
Computer Vision (CV), 
compositional generalization is 
increasingly recognized as a fundamental 
research problem~\cite{jiang-etal-2022-mutual,zhou-etal-2023-data,li-etal-2023-learning,patel2022revisitingcompositionalgeneralizationabilities,li2023exploring,11146492,li2025multisourcedcompositionalgeneralizationvisual,wiedemer2023compositional}.
Generally speaking, these studies focus on
generalizing the learned model
to out-of-distribution (OOD) scenario
from a sufficient number of compositions
in the training distribution. For example,
\cite{jiang-etal-2022-mutual} proposes
mutual exclusive training to augment the number 
of unique primitives to mitigate
compositional challenges.
\cite{li2025multisourcedcompositionalgeneralizationvisual} introduces a retrieval-augmented 
framework aimed at improving the multi-sourced compositional generalization in the field of visual question answering.
\cite{wiedemer2023compositional} formally 
defines the compositional generalization,
and provides a theorem of when a function
can be generalized.

Despite its significant advancements 
in various domains, we observe 
that research into compositional 
generalization remains underdeveloped 
in sequential recommendation systems.
In our study, we bridge this gap 
by proposing a simple multiple-choice 
auxiliary task can serve as a signal
to facilitate compositional generalization.

\section{Proposed Framework}
\label{sec:method}

\begin{figure*}[ht]
    \centering
    \includegraphics[width=0.95\linewidth]{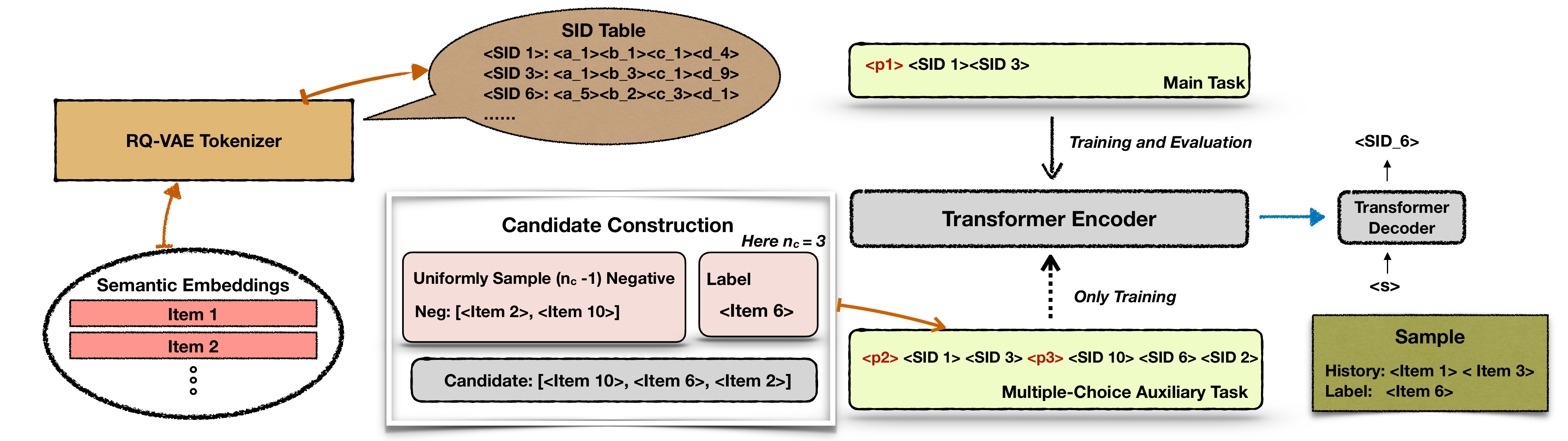}
    \caption{The overall framework of our approach
    to integrate the multiple-choice task as an
    auxiliary. In this illustration, we assume a candidate size of $n_c=3$.}
    \label{fig:whole_structure}
\end{figure*}

In this section, we formally
define the problem scope in this
work and then present our
proposed framework along with detailed
analysis.
Figure~\ref{fig:whole_structure} illustrates the overall framework. 

\subsection{Problem Formulation \& Preliminary}

\label{sec:method:prob}

In this paper, we focus on the sequential recommendation task. 
Formally, let $u \in \mathcal{U}$ denote a user and $i_{u_t} \in \mathcal{I}$ represent the target next item with which the user $u$ interacts at time $t$. Let $g(\cdot)$ be a representation function that maps the user's interaction history sequence, $s_{u_t} = [i_{u_1}, i_{u_2}, \dots, i_{u_{t-1}}] \in \mathcal{S}_u$, to a latent representation $h_{u_t} \in \mathbb{R}^{D}$. The objective of the recommendation system is to learn a well-behaved representation $h_{u_t} = g(s_{u_t})$ and a scoring function $s(h, i)$ such that 
the ground-truth next item $i_{u_t}$ is
ranked as high as possible.

For generative models~\cite{rajput2023recommendersystemsgenerativeretrieval,10.1145/3627673.3679569,10.1145/3726302.3729989}, 
SIDs are constructed from a sequence of 
semantic tokens
based on their 
semantic meaning. 
In this paper, we focus on using RQ-VAE~\cite{zeghidour2021soundstream,rajput2023recommendersystemsgenerativeretrieval} to
obtain semantic tokens:
\begin{equation}
    \text{RQ-VAE} (i_{u_t}) = (q^1_{u_t}, \, q^2_{u_t},\, \ldots, \, q^m_{u_t})\, ,
\end{equation}
where $q^l_{u_t}$ denotes the token
at level $l$.
Each item is represented by the concatenation of 
its corresponding semantic tokens to form its SID,
and the user sequence $s_{u_t}$ is then transformed
into the concatenation of the SIDs of interacted items.
The learning objective $\mathcal{L}_{\rm main}$ 
is formulated as the negative log-likelihood:
\begin{equation}
    \mathcal{L}_{\rm main} = - \sum_{\substack{s_{u_t} \in \mathcal{S}_u \\ u \in \mathcal{U}}} \sum_{l=1}^m
    \log P(q^l_{u_t} \, | \, q^{< l}_{u_t}, \, s_{u_t}) \, ,
    \label{eq:main_obobjective}
\end{equation}
where $q^{< l}_{u_t}$ denotes the semantic tokens
before $q^l_{u_t}$.
The score
function $s(h_{u_t}, i)$ is then modeled as the beam score 
in the autoregressive generation process with beam search. 

\subsection{Multiple-Choice Auxiliary Task}

\label{sec:method:mul-choice}

We design the multiple-choice auxiliary task by extending 
the representation function $g$ to additionally condition 
on the candidate set $c_{u_t}$.

\begin{equation}
g(s_{u_t}) \rightarrow g(s_{u_t}, c_{u_t}), \quad i_{u_t} \in c_{u_t} \, .
\end{equation}

The multiple-choice task is introduced as an auxiliary to the main sequential recommendation task where both tasks
share the same representation function $g$.
For the main task, the candidate set is not used, and the 
model reduces to the original formulation $g(s_{u_t})$.

\subsubsection{\textbf{Task Identifiers}}

To distinguish between
the two tasks, we introduce task identifiers 
along with a separator token.
We incorporate a 
mechanism similar to
prefix-tuning~\cite{li-liang-2021-prefix}, by prepending a learnable
header token, $p_1$ or $p_2$, in front of the user history sequence, where $p_1$ and $p_2$
denote the main task and the auxiliary task, respectively.
Also, a separating token $p_3$ is introduced to 
mark the boundary between the history sequence and the candidate set.

\begin{align}
\label{eq:inputs}
Inputs =
\begin{cases}
[p_1 \,s_{u_t}], & \text{for the main task} \\
[p_2 \,s_{u_t} \,p_3\, c_{u_t}], & \text{for the auxiliary task} \, .
\end{cases}
\end{align}
The three special tokens, $p_1$, $p_2$ and $p_3$,
possess their own independent token embeddings, and 
the models are expected to learn suitable 
representations that help distinguish the tasks
and separate the history sequence from the candidates.

\subsubsection{\textbf{Candidate Construction for Auxiliary Task}}
The candidate set $c_{u_t}$ with $n_c$ items is formed by combining the ground truth item $i_{u_t}$ and $n_c - 1$ negative
examples uniformly sampled from any unseen items as:

\begin{equation}
\begin{aligned}
c_{u_t} &= neg_{u_t} \cup \{ i_{u_t} \} \\
neg_{u_t} &\sim \mathcal{I} \setminus (s_{u_t} \cup \{ i_{u_t} \})\,.
\end{aligned}
\end{equation}
After sampling, the candidate items are randomly shuffled and then arranged in a sequential order, one item at a time, $c_{u_t} = ( c_1 c_2 \ldots c_{n_c})$\footnote{Here, $c_i$ represents the SID of the corresponding item.}.
The goal of the auxiliary task is to select
the positive item $i_{u_t}$ from the candidate set
$c_{u_t}$. The candidate set is dynamically refreshed 
for each sequence $s_{u_t}$ in each epoch. 
In a special case when $n_c=1$,
$neg_{u_t} = \emptyset$ is an empty set. Therefore,
the objective can be reinterpreted as an
answer-memorization task.

\subsubsection{\textbf{Training Objective}}
The multiple-choice auxiliary task can be
naturally formulated as sequence-to-sequence problems,
conditioned both on the historical sequence 
and the candidate set.
Therefore, we can extend the training objective in Equation~\ref{eq:main_obobjective} as:
\begin{equation}
    \mathcal{L}_{\rm aux} = - \sum_{\substack{s_{u_t} \in \mathcal{S}_u \\ u \in \mathcal{U}}}  \sum_{l=1}^m
    \log P(q^l_{u_t} \, | \, q^{< l}_{u_t}, \, s_{u_t}, \,c_{u_t}) \, .
\end{equation}
It is noteworthy that the design of the
auxiliary task shares the same decoding
targets (i.e., the ground truth next item) as 
the main task, which helps
minimize potential objective conflicts.

Combining the original main task objective 
$\mathcal{L}_{\rm main}$, 
the overall training loss $\mathcal{L}$ is the sum
of both:
\begin{equation}
    \mathcal{L} = \mathcal{L}_{\rm main} + \mathcal{L}_{\rm aux} \, .
\end{equation}
In a training epoch, each training sequence ($s_{u_t}$) is 
used once for the main task and 
once for the auxiliary task.
Each batch is then randomly constructed 
from the combined training samples, 
regardless of the task type. 

\subsection{Theoretical Analysis}

\label{sec:method:analysis}

Here we examine the 
advantages and fundamental benefits brought by the 
multiple-choice auxiliary task.

\subsubsection{\textbf{The Viewpoint of Information Theory}}
\label{sec:method:analysis:information}
First, we start to analyze 
the advantage from
the perspective of information theory.
Before integrating the auxiliary task,
the model is trained to maximize the
mutual information $I(G;I^+)$
between its latent representations
$G = g(S)$ and the target next-item $I^+$.
Here, we denote $S$ and $I^+$ as the
random variables of the historical sequence
and the target positive item, respectively.

When incorporating the candidate set 
$C$, the latent representation is 
extended to $G = g(S, C)$.
For $n_c > 1$, the candidate
set contains negative items, 
$C = \{ I^+, I^-\} = \{ I^+,\, I^-_{1}, \, \ldots,\, I^-_{n_c-1}\}$.
The model is asked to
select $I^+$ instead of all other negative items
in $I^-$.
That is,
we implicitly impose an information bottleneck~\cite{tishby2000informationbottleneckmethod} to maximize the mutual information
with $I^+$ while minimizing the mutual information
with $I^-$ as:

\begin{equation}
    max_{g} [ \, I(G;I^+ \mid C \, ) - \beta I(G;I^- \mid C \, )\, ] \, ,
    \label{eq:ib_all}
\end{equation}
where $\beta$ is an implicit Lagrange multiplier.
Equation~\ref{eq:ib_all} aligns with
the philosophy of 
BPR~\cite{rendle2012bprbayesianpersonalizedranking} 
ranking, as the latent representation is learned
to prefer positive item to negative items.
Since we still preserve the main
task, whose target is to maximize $I(G;I^+)$,
we may further assume that the model will also converge
to maximize $I(G;I^+ \mid C)$. In this case,
we naturally have

\begin{equation}
    \beta I(G;I^- \mid C \,)  \ll I(G;I^+ \mid C \,) \, ,
    \label{eq:ib_small}
\end{equation}
as the model starts to converge.

From Equation~\ref{eq:ib_small}, 
we can regard negative
items as small noise perturbation. 
To remain consistent predictions across
different negative items, 
$g$ should be robust.
This perspective is particularly useful in recommendation, where the user sequence
may contain items that 
do not perfectly match the user’s current intent.

\subsubsection{\textbf{Facilitating Compositional Generalization}}
\label{sec:method:analysis:composition}

Beyond the perspective of information
theory, incorporating candidate items as
inputs introduces a natural structure
that facilitates compositional
generalization.

In the generative retrieval process, 
the training distribution of the main task 
is highly sparse, where many item compositions
rarely or never co-occur.
Consequently, the model struggles to capture the underlying behaviors of unseen compositions.
For instance, users interested in guitars are 
more likely to purchase relevant accessories rather than items related to orchestra.
This lack of joint
observations for compositions
between guitars and orchestra could lead to weak generalization on such unseen combinations during testing. 
The situation can potentially be even worse,
when items share 
similar semantic information but exhibit 
different behavioral patterns. The 
ill-defined representation function $g$
may induce 
unintended correlations. 
Since the training distribution 
lacks the composition for these 
unrelated items, predefined semantics force 
them to share a similar representation,
leading to incorrect recommendations.

The multiple-choice task then augments the training 
distribution by providing candidate items, and asking the model to pick the correct one. 
Through this process, the model starts to capture
the compositions between items driven by different intents,
which facilitates the learning of a more well-defined representation function to generalize from the training distribution to the test distribution.
In particular, since $i_{u_t} \in c_{u_t}$, the auxiliary
task forces the model to memorize the
positive signal between history
$s_{u_t}$ and target $i_{u_t}$.
This is crucial because 
the training labels could
become part of the input sequence.
While the previous sequential task
constructions
often overlook
these positive compositions,
the multiple-choice task can also effectively
bridge the gap.

\section{Experiments}

\label{sec:exp}

In this section, we present our experimental
setup and results on four real-world datasets.
The experiments are designed to answer
the following research questions ($\ba{RQs}$):

\begin{itemize}[leftmargin=*]
    \item \ba{RQ1:} How does the multiple-choice auxiliary 
    task affect the performance compared to standard baselines,
    and how does it vary 
    across different values of $n_c$?
    \item \ba{RQ2:} How does the model perform across various scenarios, and what key factors drive its behavior?
    \item \ba{RQ3:} What is the relative importance of the two tasks 
    during the training process, and what is the impact of different 
    candidate sampling methods?
    \item \ba{RQ4:} How does the multiple-choice task behave in our setting?
\end{itemize}

\subsection{Experimental Settings}

\label{sec:exp:data-and-model}

\subsubsection{\textbf{Datasets}}
We consider four real-world datasets across various kinds of sequential recommendation scenarios.
\begin{itemize}[leftmargin=*]
    
\item 
\bi{Amazon Industrial and Scientific} and \bi{Musical Instruments} are from the Amazon'23 review dataset~\cite{hou2024bridging}, 
including interactions ranging from 1996 to 2023. 
This dataset is notable for its high sparsity. For simplicity, we refer to these two datasets as \bi{Scientific} and \bi{Instruments}, respectively.

\item 
\bi{Yelp} \footnote{\url{https://www.yelp.com/dataset}} 
is a widely used dataset for business recommendation. We adopt the item metadata 
and interaction sequences as preprocessed
by LETTER~\cite{10.1145/3627673.3679569} 
for our experiments.

\item
\bi{MovieLens1M}~\cite{10.1145/2827872} 
is the 1-million-interaction dataset derived from user interactions with the \textit{MovieLens} online recommender system. The user-item interactions are denser compared to Amazon and Yelp datasets.

\end{itemize}

For preprocessing, following previous works~\cite{zheng2024adaptinglargelanguagemodels,10.1145/3627673.3679569},
standard 5-core filtering and leave-one-out strategy
are applied. 
Table~\ref{tab:dataset-stats} summarizes the statistics of each dataset. 

\begin{table}[!t]
    \centering
    \caption{Dataset statistics for the four real-world benchmarks.}
    \label{tab:dataset-stats}
        \begin{small}
    \begin{tabular}{lccccc}
        \toprule
        Dataset & \# Users & \# Items & \multicolumn{2}{c}{\shortstack{History Length \\ per user}} & Sparsity \\
        \cmidrule(r){4-5}
        & & & Mean & Median & \\
        \midrule
        Scientific              & 50,985 & 25,848 & 8.10 & 6 & 99.97 \% \\
        Instruments & 57,439 & 24,587 & 8.91 & 7 & 99.96 \% \\
        Yelp      & 30,431 & 20,033 & 10.40 & 7 & 99.95 \% \\
        MovieLens 1M      & 6,040 & 3,416 & 165.50 & 96 & 95.16 \% \\
        \bottomrule
    \end{tabular}
        \end{small}
\end{table}

\subsubsection{\textbf{Implementation Details}}
\label{sec:exp:data-and-model:model-imple}

To derive SIDs, we apply pretrained Sentence-T5~\cite{ni2022sentence} (\texttt{sentence-t5-xl}) to generate semantic embeddings for RQ-VAE.
The layer number of RQ-VAE codes is set to $4$, with codebook size in each level being $256$, across all datasets.
The three special tokens $p_1$, $p_2$ and $p_3$ are introduced as three independent new tokens.
The
learning rate is set to $5 \times 10^{-4}$,
and we set the
maximum history length to $20$, following previous
generative recommendation models~\cite{zheng2024adaptinglargelanguagemodels,10.1145/3627673.3679569}. The number of epochs is set to $200$, with early 
stopping patience of $20$ based only on the main task 
validation loss.
During generation, constrained beam search is applied with beam size $20$.

\subsubsection{\textbf{Baseline Methods}}
\label{sec:exp:data-and-model:other-base}

We compare our framework with two categories of conventional baseline methods as follows:

\begin{itemize}[leftmargin=14pt]
    \item[(1)] Traditional Baselines:
    \begin{itemize}[leftmargin=*]
        \item \ba{Caser}~\cite{tang2018personalized} applies
    horizontal and vertical convolution layers
    to capture union-level and point-level sequential
    patterns.
        \item \ba{Bert4Rec}~\cite{sun2019bert4recsequentialrecommendationbidirectional} performs sequential recommendation
    on a bidirectional transformer architecture with mask cloze
    task as its objective.
        \item \ba{GRU4Rec}~\cite{hidasi2015session} uses RNN to capture
    sequential patterns.
        \item \ba{SASRec}~\cite{kang2018selfattentivesequentialrecommendation}
    utilizes self-attention mechanism to model sequences and
    predict the next item.
        \item \ba{HGN}~\cite{ma2019hierarchical} applies a hierarchical gating network to capture both the user long-term and short-term
    interests.
    \end{itemize}
    \item[(2)] Generative Baselines:
    \begin{itemize}[leftmargin=*]
        \item \ba{TIGER}~\cite{rajput2023recommendersystemsgenerativeretrieval} uses text embedding from metadata information to construct SIDs via RQ-VAE, and generates the next item via an encoder-decoder architecture. 
            We consider three variants with different SID constructions. 
            \ba{TIGER} and \ba{TIGER-SAS} utilize pretrained  Sentence-T5~\cite{ni2022sentence} 
            embeddings and finetuned embeddings learned from SASRec~\cite{kang2018selfattentivesequentialrecommendation} to derive SIDs.
            \ba{LETTER-TIGER}~\cite{10.1145/3627673.3679569} integrates collaborative and diversity regularization to learn the SID tokenizer.
        \item \ba{ETEGRec}~\cite{10.1145/3726302.3729989} integrates the item tokenizer and generative retrieval into an end-to-end framework, based on the approach of recommendation-oriented alignment.
    \end{itemize}
\end{itemize}

For traditional baselines, we adopt the open-source recommendation framework RecBole~\cite{zhao2021recbole,zhao2022recbole} to implement all methods.
RecBole adopts cross-entropy loss for these
baseline methods, leading to stronger and more competitive baselines.
For generative baselines, TIGER and TIGER-SAS follow the same 
SID construction method and implementation setup as described in Section~\ref{sec:exp:data-and-model:model-imple} for a fair comparison.\footnote{To clarify, they don’t contain the three newly introduced tokens, $p_1$, $p_2$ and $p_3$.}
For LETTER-TIGER~\cite{10.1145/3627673.3679569}, we use LLaMA-2-7B~\cite{touvron2023llama} to construct the semantic embeddings, while
other details follow the official implementation.
For ETEGRec, we obtain the SASRec semantic embeddings using RecBole and subsequently follow the official implementation.

\subsection{Overall Performance (RQ1)}

\label{sec:exp:overall-performance}
In this section, we evaluate the overall performance
of our approach and baselines to investigate the effectiveness of the auxiliary task to answer RQ1.

\begin{table*}[ht]
    \centering
    \caption{Overall Performance Comparison. We
    denote our approach as ``TIGER+$n_c$'', where
    $n_c$ is the number of candidates.
    Bold and \bu{underlined font} are
    used to denote the best and second-best results.
    * denotes a significant improvement ($p<0.05$) compared
    to TIGER via paired t-test on per-user metrics.
    }
    \label{tab:main-with-baseline}
    \vspace{-6pt}
    \begin{small}
    \setlength{\tabcolsep}{2.5pt}
    \begin{tabular}{l cccc cccc cccc cccc}
        \toprule
        \multirow{2}{*}{\textbf{Methods}} 
        & \multicolumn{4}{c}{Scientific} 
        & \multicolumn{4}{c}{Instruments}
        & \multicolumn{4}{c}{Yelp} 
        & \multicolumn{4}{c}{MovieLens} \\
        \cmidrule(lr){2-5} 
        \cmidrule(lr){6-9}
        \cmidrule(lr){10-13}
        \cmidrule(lr){14-17}
        & H@5 & N@5 & H@10 & N@10 
        & H@5 & N@5 & H@10 & N@10
        & H@5 & N@5 & H@10 & N@10
        & H@5 & N@5 & H@10 & N@10
        \\
        \midrule
        Caser
        & 0.0129 & 0.0082 & 0.0218 & 0.0110
        & 0.0216 & 0.0134 & 0.0360 & 0.0180
        & 0.0291 & 0.0208 & 0.0403 & 0.0244
        & 0.1354 & 0.0873 & 0.2123 & 0.1120 \\
        Bert4Rec
        & 0.0144 & 0.0092 & 0.0239 & 0.0122
        & 0.0257 & 0.0164 & 0.0407 & 0.0212
        & 0.0167 & 0.0105 & 0.0278 & 0.0141
        & 0.1535 & 0.0988 & 0.2449 & 0.1283 \\
        GRU4Rec
        & 0.0199 & 0.0127 & 0.0319 & 0.0166
        & 0.0301 & 0.0193 & 0.0475 & 0.0250
        & 0.0225 & 0.0144 & 0.0384 & 0.0195
        & 0.1695 & 0.1124 & 0.2515 & 0.1386 \\
        SASRec
        & 0.0250 & 0.0147 & 0.0391 & 0.0193
        & 0.0335 & 0.0210 & 0.0526 & 0.0272
        & \bu{0.0435} & \ba{0.0325} & \bu{0.0613} & \ba{0.0383}
        & 0.1430 & 0.0898 & 0.2262 & 0.1164 \\
        HGN
        & 0.0207 & 0.0130 & 0.0346 & 0.0174
        & 0.0305 & 0.0195 & 0.0484 & 0.0252
        & 0.0367 & 0.0253 & 0.0540 & 0.0308
        & 0.0758 & 0.0481 & 0.1280 & 0.0649 \\
        \midrule
        TIGER 
        & 0.0255 & 0.0167 & 0.0398 & 0.0213
        & 0.0353 & 0.0228 & 0.0554 & 0.0293
        & 0.0243 & 0.0158 & 0.0390 & 0.0205
        & 0.1811 & 0.1258 & 0.2627 & 0.1520 \\
        TIGER-SAS
        & 0.0227 & 0.0148 & 0.0363 & 0.0192
        & 0.0335 & 0.0222 & 0.0523 & 0.0282
        & 0.0256 & 0.0161 & 0.0441 & 0.0220
        & 0.1909 & \bu{0.1330} & 0.2743 & \ba{0.1598} \\
        LETTER-TIGER
        & 0.0255 & 0.0164 & 0.0401 & 0.0211
        & 0.0354 & 0.0230 & 0.0555 & 0.0294
        & 0.0289 & 0.0186 & 0.0466 & 0.0243
        & 0.1934 & 0.1326 & \bu{0.2778} & \bu{0.1597} \\
        ETEGRec
        & 0.0269 & 0.0172 & 0.0425 & 0.0222
        & \ba{0.0385} & 0.0250 & \ba{0.0598} & \bu{0.0318}
        & 0.0336 & 0.0214 & 0.0540 & 0.0279
        & \ba{0.1959} & 0.1325 & \ba{0.2790} & 0.1594 \\
        \midrule
        TIGER+$n_c=2$
        & \bs{\bu{0.0274}} & \bs{\bu{0.0179}} & \bs{\bu{0.0426}} & \bs{\bu{0.0228}}
        & \bs{\ba{0.0385}} & \bs{\bu{0.0253}} & \bs{0.0587} & \bs{\bu{0.0318}}
        & \bs{0.0320} & \bs{0.0225} & \bs{0.0489} & \bs{0.0279}
        & \bs{\bu{0.1950}} & \bs{\ba{0.1338}} & \bs{0.2742} & \bs{0.1594} \\
         TIGER+$n_c=10$
        & \bs{\ba{0.0284}} & \bs{\ba{0.0184}} & \bs{\ba{0.0442}}  & \bs{\ba{0.0235}} 
        &\bs{\ba{0.0385}} & \bs{\ba{0.0256}} & \bs{\bu{0.0594}}  & \bs{\ba{0.0323}} 
        & \bs{\ba{0.0444}} & \bs{\bu{0.0310}} & \bs{\ba{0.0621}} & \bs{\bu{0.0367}}
        & \bs{0.1921} & \bs{0.1314} & \bs{0.2770} & \bs{0.1586} \\
        \midrule
         \itt{Gain (v.s TIGER)} 
         & \itt{11.4 \%} & \itt{10.2 \%} & \itt{11.1 \%} & \itt{10.3 \%}
         & \itt{9.1 \%}  & \itt{12.3 \%} & \itt{7.2 \%}  & \itt{10.2 \%} 
         & \itt{82.7 \%} & \itt{96.2 \%} & \itt{59.2 \%} & \itt{79.0 \%} 
         & \itt{7.7 \%}  & \itt{6.4 \%}  & \itt{5.4  \%} & \itt{4.9 \%}
        \\
        \bottomrule
    \end{tabular}
    \end{small}
\end{table*}

We first evaluate the performance 
differences against baselines.
We consider two settings: 
one with a single negative and 
one positive item (TIGER+$n_c=2$), 
and the other with 10 candidate items (TIGER+$n_c=10$).
We use
common recommendation metrics, Hit@K and NDCG@K, 
abbreviated as H@K and N@K, with
$K \in \{5,10\}$. The results are reported in
Table~\ref{tab:main-with-baseline}.
Detailed comparison with multiple runs across different seeds are shown in Table~\ref{tab:metrics_multiple_seed} in Appendix~\ref{sec:append:more-exp}.

From Table~\ref{tab:main-with-baseline}, we observe
that incorporating the multiple-choice
task as an auxiliary objective
yields significant performance
improvements over the base model TIGER
across all datasets. 
Additionally, the performance is competitive with
mainstream generative
models, even achieving the
best ranking among all
baselines on the
\textit{Scientific} datasets.
Except for \textit{MovieLens}, ``TIGER+$n_c=10$''
consistently ranks among the 
best or second-best methods.
Notably, our approach remains 
simple, interpretable,
and does not rely on extensive hyperparameter 
tuning.
We attribute this improvement to
a more robust
(Section~\ref{sec:method:analysis:information}) 
and 
well-defined (Section~\ref{sec:method:analysis:composition})
representation
$g$, resulting from the integration 
of the multiple-choice auxiliary task.

\begin{figure}
    \centering
    \includegraphics[width=0.49\linewidth]{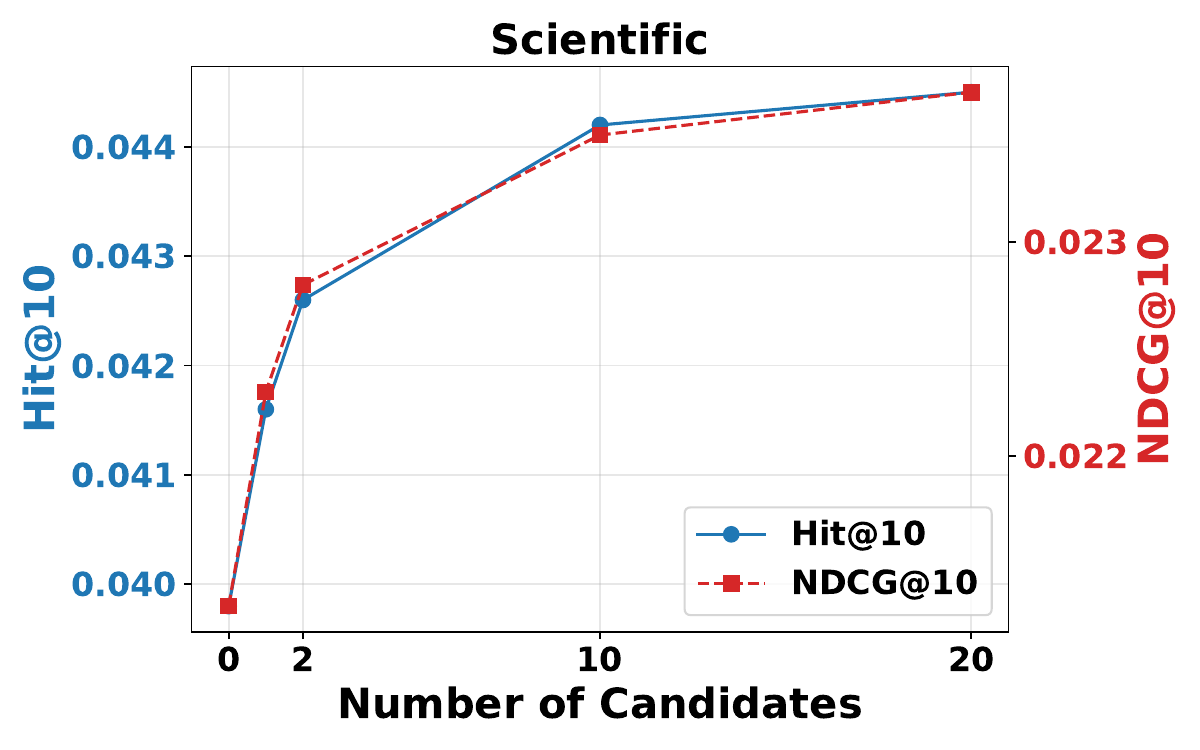}
    \includegraphics[width=0.49\linewidth]{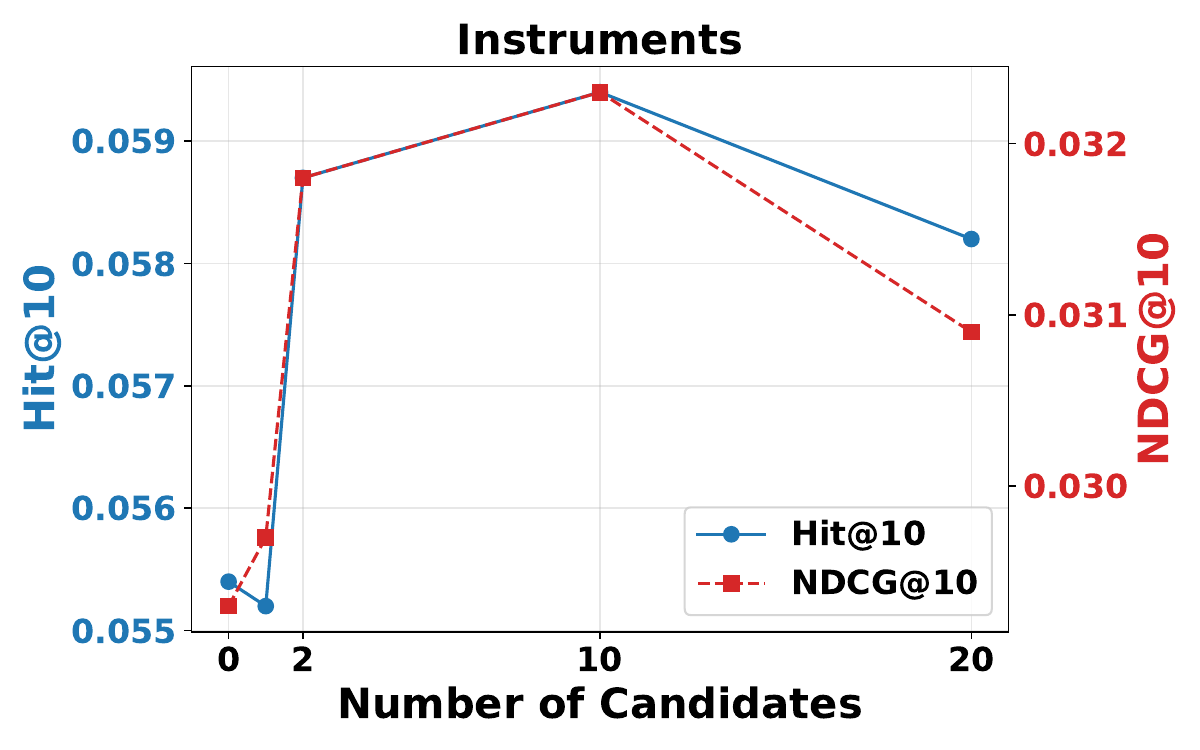} \\
    \includegraphics[width=0.49\linewidth]{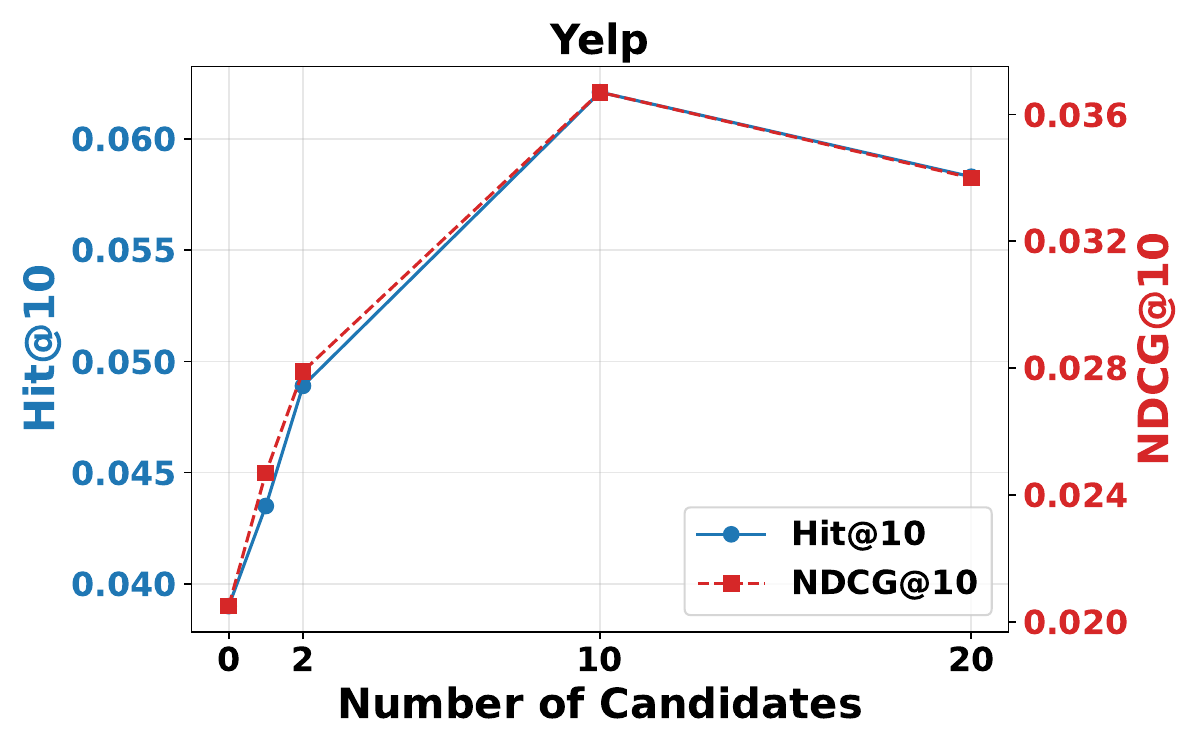}
    \includegraphics[width=0.49\linewidth]{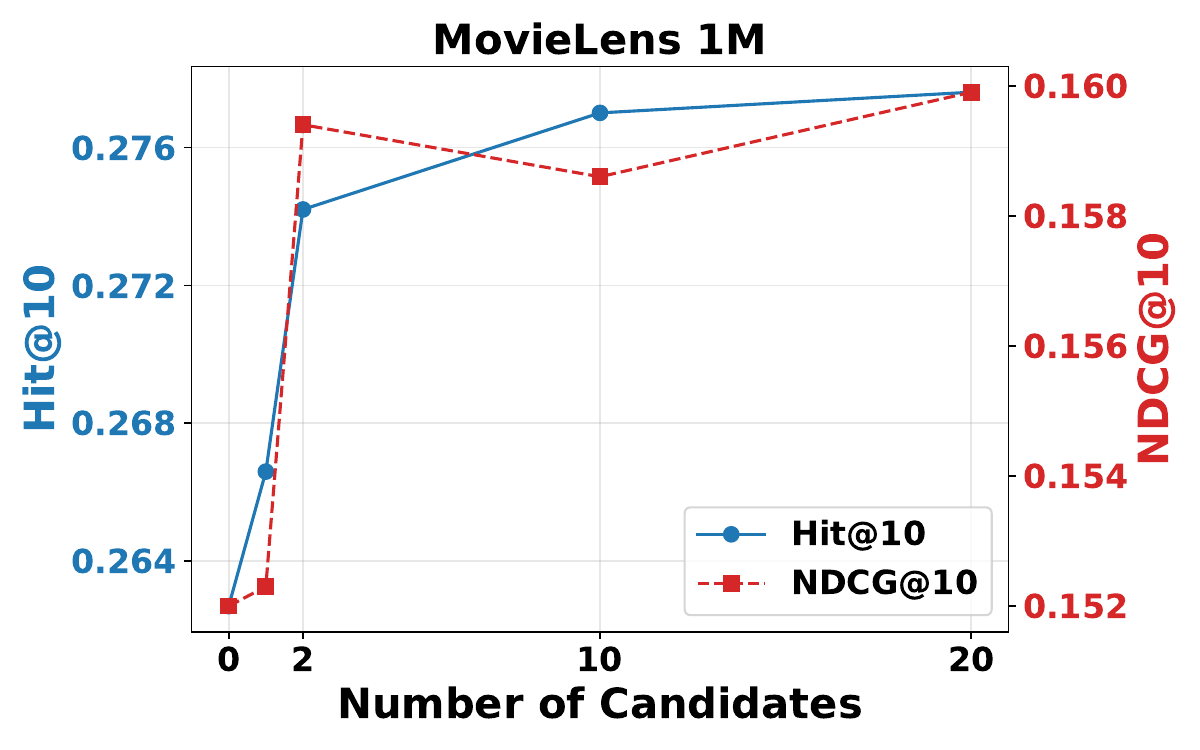}
    \caption{Performance (Hit@10 and NDCG@10) v.s Number of Candidates ($n_c$). Here, $n_c = 0$ indicates the
    performance of base model TIGER specifically.}
    \label{fig:num_performance}
    \vspace{-9pt}
\end{figure}

\begin{figure*}[!t]
    \centering
    \includegraphics[width=0.91\textwidth]{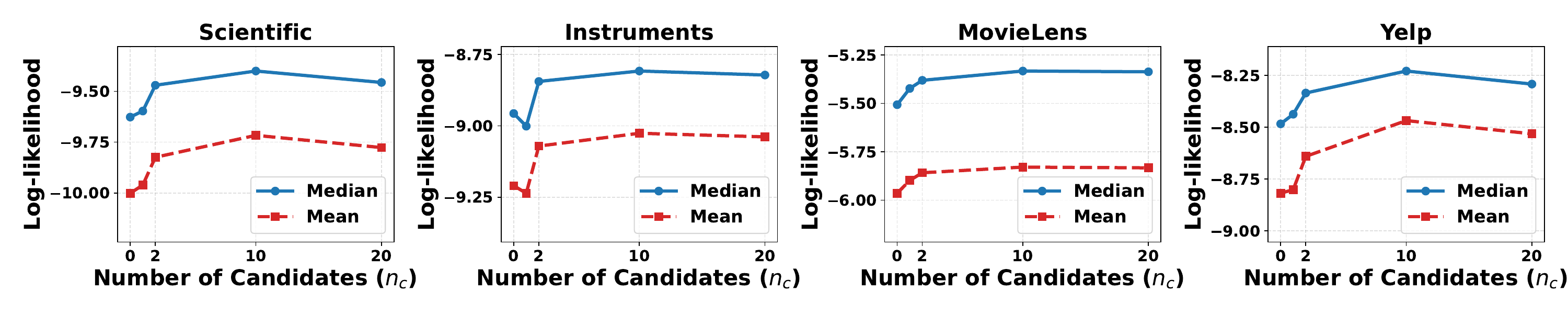} 
    \caption{Log likelihood vs Number of Candidates ($n_c$) across different datasets. 
    The log-likelihood is computed via
    the teacher forcing decoding. We aggregate the
    values from the first semantic token
    to EOS. 
    Here, $n_c = 0$ means
    the base model TIGER specifically.}
    \label{fig:loglike}
\end{figure*}

\noindent\textbf{Impact of Candidate Number.}
Subsequently, we conduct the performance study across 
different values of $n_c$. 
As shown in Figure~\ref{fig:num_performance},
we observe a trend of performance improvement as $n_c$ 
increases. When $n_c$ becomes excessively large, the performance begins to saturate. This suggests that the model may have already captured sufficient compositional patterns among items. 
Additional
negative signals could become redundant, and
a larger number of candidates may increase
the likelihood of including false
negatives, defined as items unseen by the users, but aligned with their intents, thus resulting in a
performance drop.

In particular, when $n_c = 1$, we still observe improvements on several 
datasets. Although the gains are generally 
less significant, this 
suggests that explicitly learning compositional patterns between the training label $i_{u_t}$ and its corresponding sequence $s_{u_t}$ can already benefit the models at test time.

\subsection{In-depth Analysis and Discussion}
\label{sec:exp:ablation}

\subsubsection{\textbf{Model Confidence on Ground-Truth Labels} (RQ2)}

To understand how confidently our models perform on all testing samples, 
we start the in-depth analysis by
probing into its likelihood of
generating ground-truth next items. 
As shown in Figure~\ref{fig:loglike},
when $n_c$ increases,
we observe
an overall improvement of median likelihood values
across different datasets.
As $n_c$ continues to grow, the likelihood values
eventually begin to saturate.
The growth trends align with the performance 
observed earlier (Figure~\ref{fig:num_performance}).
The results suggest that
the model develops a higher confidence 
regarding the ground-truth items in the majority of
testing samples.

\subsubsection{\textbf{Predicting Items with Different Frequency} (RQ2)}
\label{sec:exp:ablation:long-tail}
Next, we examine the performance
across items with different interaction
frequency. 
Specifically,
we evaluate on three test subsets,
``popular,'' ``medium,'' and ``long-tail'',
where
the ground-truth items ($i_{u_t}$)
have a training interaction frequency 
that falls within the top, middle and bottom one-third
among all items. The data statistics are documented in Table~\ref{tab:stats-of-frequency} in Appendix~\ref{sec:append:more-exp}.

The evaluation results are shown in Table~\ref{tab:long-tail}. Incorporating the 
multiple-choice auxiliary task consistently improves performance 
in popular settings, while causing no significant degradation 
in medium and long-tail subsets. This suggests that 
the auxiliary task enhances the model’s discriminative ability to 
distinguish easier ground-truth items 
from other negative samples, without 
confusing harder cases. Even on \textit{Yelp}
and \textit{MovieLens}, we observe
performance improvements in the medium and
long-tail subsets. For \textit{Scientific}
and \textit{Instruments}, the performance
gains are more modest. This could be due to
a greater difficulty of the 
medium and long-tail scenarios in 
these datasets.

\begin{table}[!t]
    \centering
    \caption{Performance under different sequence splits based on the popularity of ground-truth items. The popular, medium, and long-tail splits are categorized
    according to whether the ground-truth next item’s interaction frequency in the training set falls into the top, middle, or bottom one-third of all items.
    }
    \label{tab:long-tail}
    \begin{small}
    \begin{tabular}{cc cccc}
        \toprule
        \multirow{2}{*}{\textbf{Datasets}} &
        \multirow{2}{*}{\textbf{Splits}} &
        \multicolumn{2}{c}{TIGER} & 
        \multicolumn{2}{c}{TIGER + $n_c=10$}\\
        \cmidrule(lr){3-4}
        \cmidrule(lr){5-6}
        & & H@10 & N@10  & H@10 & N@10 
        \\
        \midrule
        \multirow{3}{*}{Scientific} & popular
        & 0.0671 & 0.0359 & 0.0741 & 0.0396
        \\
        & medium
        & 0.0025 & 0.0010 & 0.0037 & 0.0016
        \\
        & long-tail
        & 0.0004 & 0.0002 & 0.0004 & 0.0001
        \\
        \midrule
        \multirow{3}{*}{Instruments} & popular
        & 0.0796 & 0.0421 & 0.0855 & 0.0466
        \\
        & medium
        & 0.0028 & 0.0011 & 0.0024 & 0.0012
        \\
        & long-tail
        & 0.0007 & 0.0003 & 0.0009 & 0.0004 \\
        \midrule
        \multirow{3}{*}{Yelp} 
        & popular
        & 0.0613 & 0.0319 & 0.0768 & 0.0420
        \\
        & medium
        & 0.0070 & 0.0041 & 0.0354 & 0.0245
        \\
        & long-tail
        & 0.0038 & 0.0023 & 0.0432 & 0.0317 \\
        \midrule
        \multirow{3}{*}{MovieLens} & popular
        & 0.3043 & 0.1794 & 0.3209 & 0.1866
        \\
        & medium
        & 0.1678 & 0.0855 & 0.1770 & 0.0908
        \\
        & long-tail
        & 0.0340 & 0.0155 & 0.0340 & 0.0194
        \\
        \bottomrule
    \end{tabular}
    \end{small}
\end{table}

\subsubsection{\textbf{Performance on Different Sequence Length} (RQ2)}

\begin{figure}
    \centering
    \includegraphics[width=0.49\linewidth]{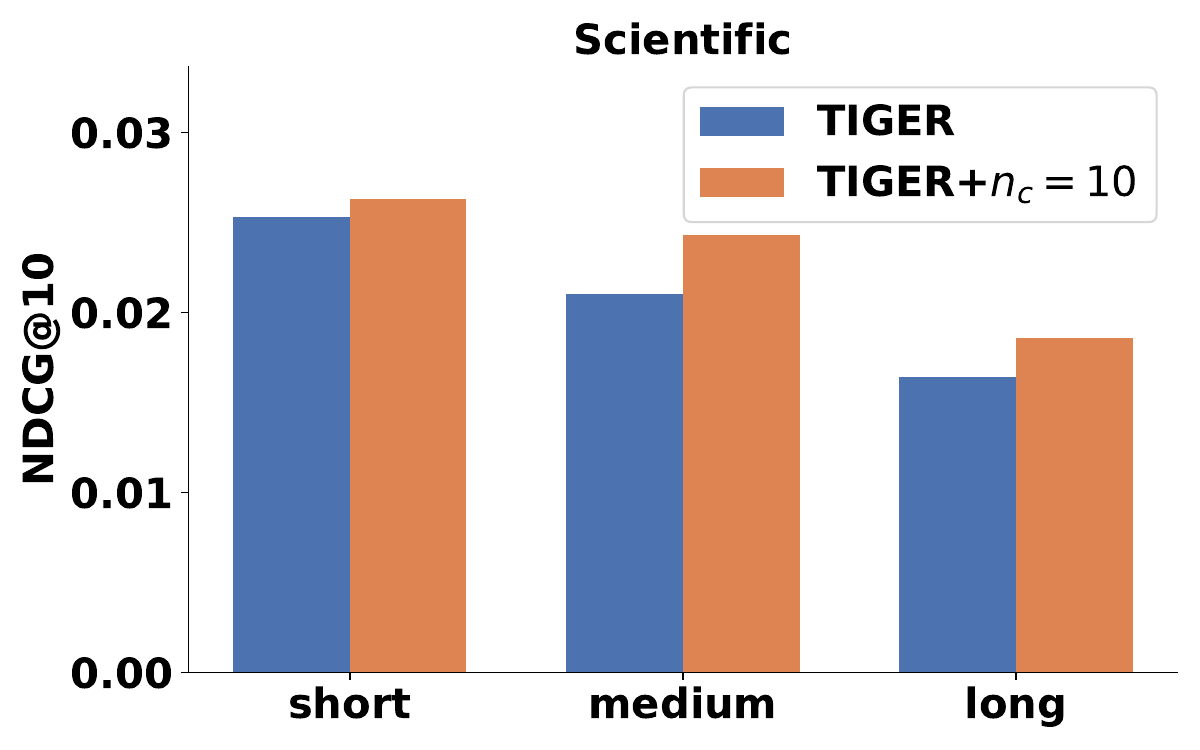}
    \includegraphics[width=0.49\linewidth]{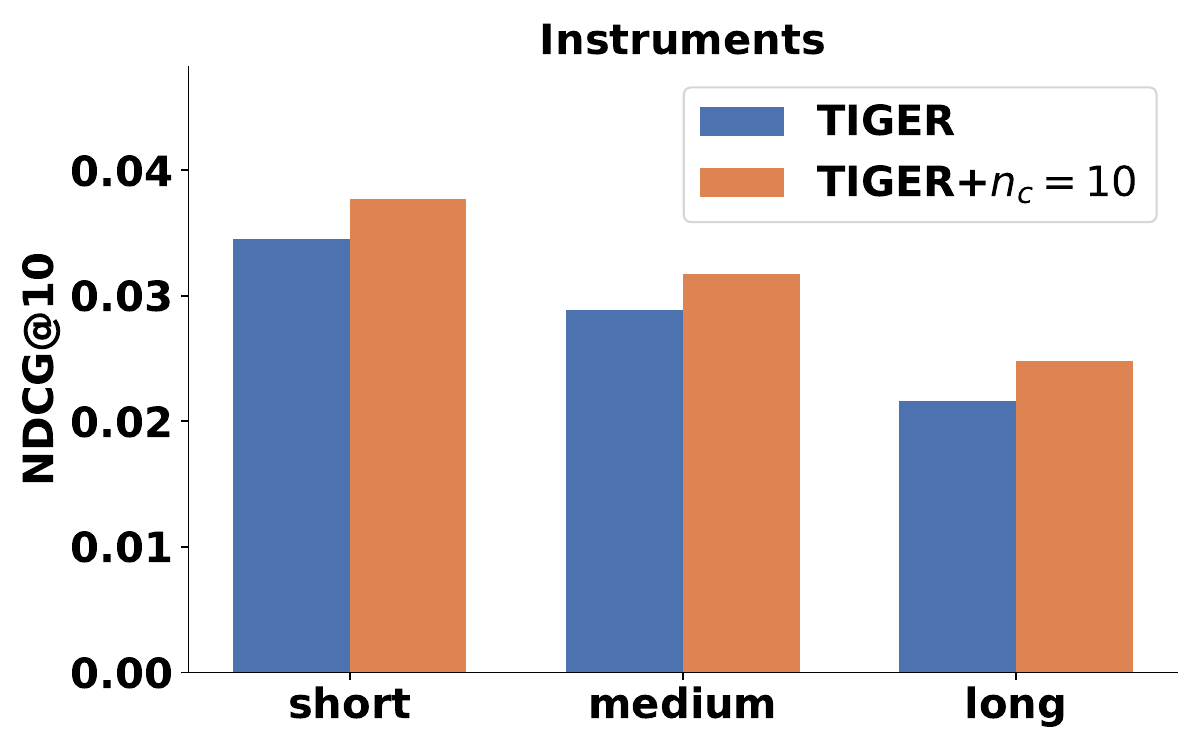} \\
    \includegraphics[width=0.49\linewidth]{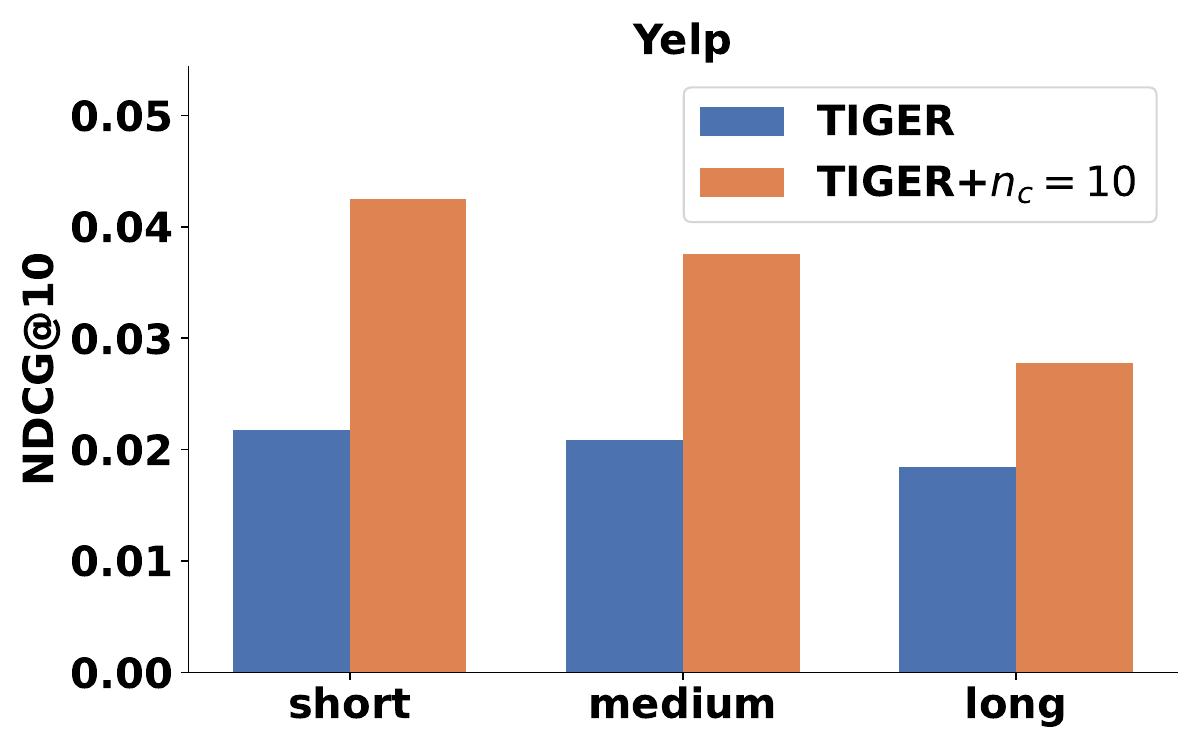}
    \includegraphics[width=0.49\linewidth]{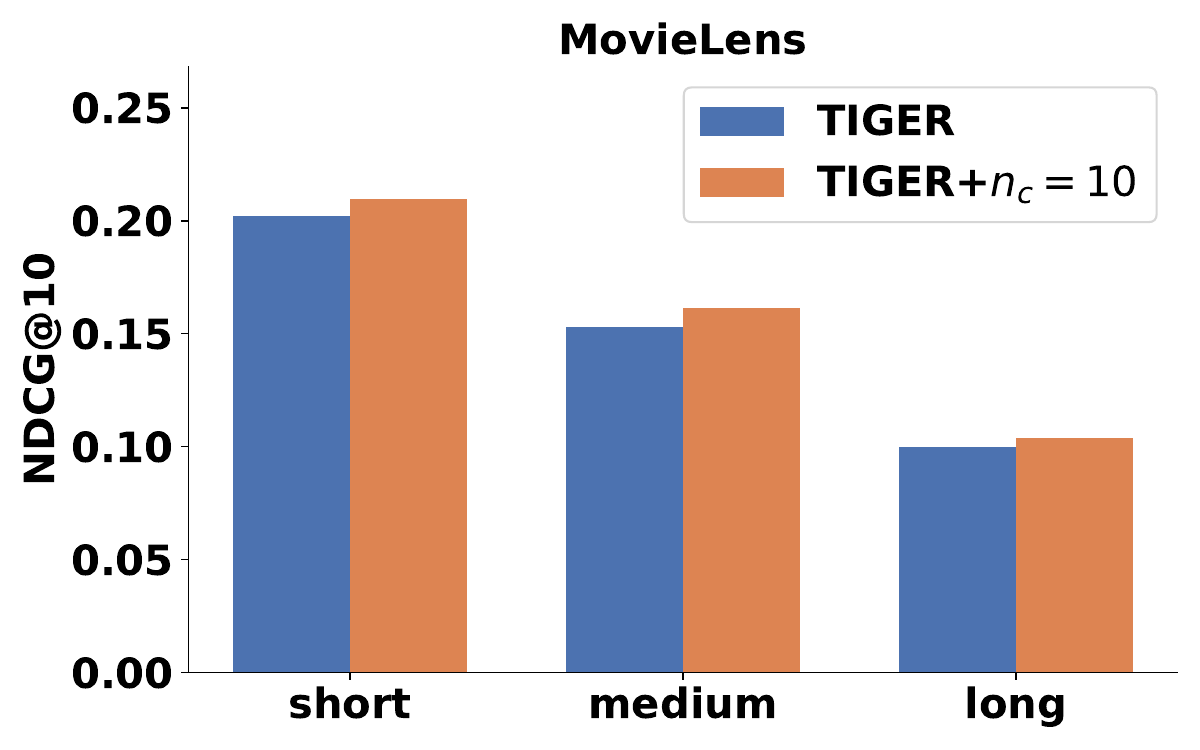} \\
    \caption{Performance difference across user sequence lengths. ``Short'', ``Medium'', and ``Long'' correspond to users whose sequence lengths fall in the bottom third, middle third, and top third in the dataset, respectively. NDCG@10 is reported.}
    \label{fig:different-seq-length}
\end{figure}

In this section, 
we further examine the performance difference across 
users with different sequence lengths. As shown in 
Figure~\ref{fig:different-seq-length}, we partition user 
sequences into short, medium, and long categories by dividing them into three bins based on sequence length, and 
compare their performance against the base model TIGER.
Among all datasets, 
incorporating the candidate set benefits performance 
for users with different sequence lengths, without exhibiting a 
clear bias toward any specific group.

From Figure~\ref{fig:different-seq-length}, we also observe 
that the performance consistently 
declines as the sequence length increases for both TIGER and ``TIGER+$n_c=10$''. This is possibly driven by the increasing 
complexity and diversity of user intents in longer interaction 
histories.
Notably, our framework still achieves the same degree of
improvement over TIGER in these challenging settings,
suggesting its robustness to long sequence scenarios.

\subsubsection{\textbf{Illustrative Examples} (RQ2)}

{\def\arraystretch{1.2}
\begin{table}[ht]
    \centering
    \begin{footnotesize}
    \begin{tabular}{l p{5cm}}
    \toprule

    \multirow{5}{*}{Interaction History}   
    & 1. Dangerous Liaisons (Drama, Romance) \\
    & 2. Babe (Children's, Comedy, Drama) \\
    & 3. Trainspotting (Drama) \\
    & 4. Color Purple (Drama) \\
    & 5. Papillon (Drama) \\
    \midrule

    \multirow{3}{*}{TIGER}    &  
    (1) Rain Man (Drama) \\
    & (2) Apollo 13 (Drama) \\
    & (3) Field of Dreams (Drama) \\

    \midrule

    \multirow{4}{*}{TIGER + $n_c=10$} & 
    \textbf{(1) E.T. the Extra-Terrestrial (Children's, Drama, Fantasy, Sci-Fi)} \\
    & (2) Apollo 13 (Drama) \\
    & (3) It's a Wonderful Life (Drama) \\

    \bottomrule
    \end{tabular}

    \caption{An example of a correct prediction in our approach from
    MovieLens dataset. The interaction history is presented sequentially, with only the most recent five items displayed. We present the top-3 predictions for this sequence, where the ground-truth item is highlighted in bold. }
    \label{tab:case_study_ml-1m}
    \end{footnotesize}
\end{table}
}

{\def\arraystretch{1.2}
\begin{table*}[ht]
    \centering
    \begin{footnotesize}
    \begin{tabular}{l p{13.5cm}}
    \toprule
    \multirow{4}{*}{Interaction History}  & 
    1. Yamaha YPG-235 76-Key Portable Grand Piano \\
    & 2. Yamaha OEM PKBB1 Adjustable Padded Keyboard X-Style Bench, Black, 19.5 inches \\
    & 3. On-Stage KS7191 Classic Double-X Keyboard Stand \\
    & 4. LyxPro HAS-30 Closed Back Over-Ear Professional Recording Headphones for Studio Monitoring, DJ and Home Entertainment,Black  \\
    \midrule
    \multirow{3}{*}{TIGER}    &  
    (1) Yamaha OEM PKBB1 Adjustable Padded Keyboard X-Style Bench, Black, 19.5 inches \\
    & (2) M-Audio SP-2 - Universal Sustain Pedal with Piano Style Action For MIDI Keyboards, Digital Pianos \& More \\
    & (3) RockJam Xfinity Heavy-Duty, Double-X, Pre-          Assembled, Infinitely Adjustable Piano Keyboard Stand with Locking Straps \\
    \midrule
    \multirow{4}{*}{TIGER + $n_c=10$} & 
    \ba{(1) Blue Yeti USB Microphone for PC, Mac, Gaming, Recording, Streaming, Podcasting, Studio and Computer Condenser Mic with Blue VO!CE effects, 4 Pickup Patterns, Plug and Play – Blackout}
    \\
    & (2) Enabled Dragonpad USA- 6" Microphone Studio Pop Filter with Clamp - BLK/WHT
    \\
    & (3) M-Audio SP-2 - Universal Sustain Pedal with Piano Style Action For MIDI Keyboards, Digital Pianos \& More 
    \\
    \bottomrule
    \end{tabular}
    \caption{An example of a correct prediction in our approach from
    Instruments dataset. The interaction history is presented sequentially. We present the top-3 predictions for this sequence, where the ground-truth item is highlighted in bold. }
    \label{tab:case_study}
    \end{footnotesize}
\end{table*}
}

Apart from the overall performance comparison, 
we further present two illustrative examples to provide qualitative insights into
how real-world datasets benefit
from the auxiliary task, shown in Table~\ref{tab:case_study_ml-1m}
and Table~\ref{tab:case_study}.

In Table~\ref{tab:case_study_ml-1m}, 
the user has watched movies that are mostly 
associated with the \textit{Drama} genre. 
As a result, TIGER produces three 
recommendations that are purely 
within the \textit{Drama} category. 
However, since a \textit{Children’s} movie 
appears in the recent 
interaction history, 
our approach captures this 
signal in the user intent 
and additionally recommends 
a movie related to both the \textit{Children’s} and
\textit{Drama} categories.

Another example is shown 
in Table~\ref{tab:case_study}, where
the user interaction history consists of
a piano, a keyboard, a keyboard stand,
and recording headphones.
Based on
the sequence, original TIGER chooses to recommend
keyboard-related accessories, which are
semantically similar to the observed history. 
In contrast, 
TIGER with the auxiliary task predicts microphone
and its accessory as top-2 choices,
which are absent from the interaction history,
but consistent with the user intent.

These two examples suggest that the auxiliary task
helps the model capture
a higher level behavior intent, instead of 
relying solely on metadata semantic similarity between items.

\subsubsection{\textbf{Relative Importance of the Two Tasks} (RQ3)}
\label{sec:exp:ablation:diff-proportion}

\begin{figure*}[!t]
    \centering
    \includegraphics[width=0.32\linewidth]{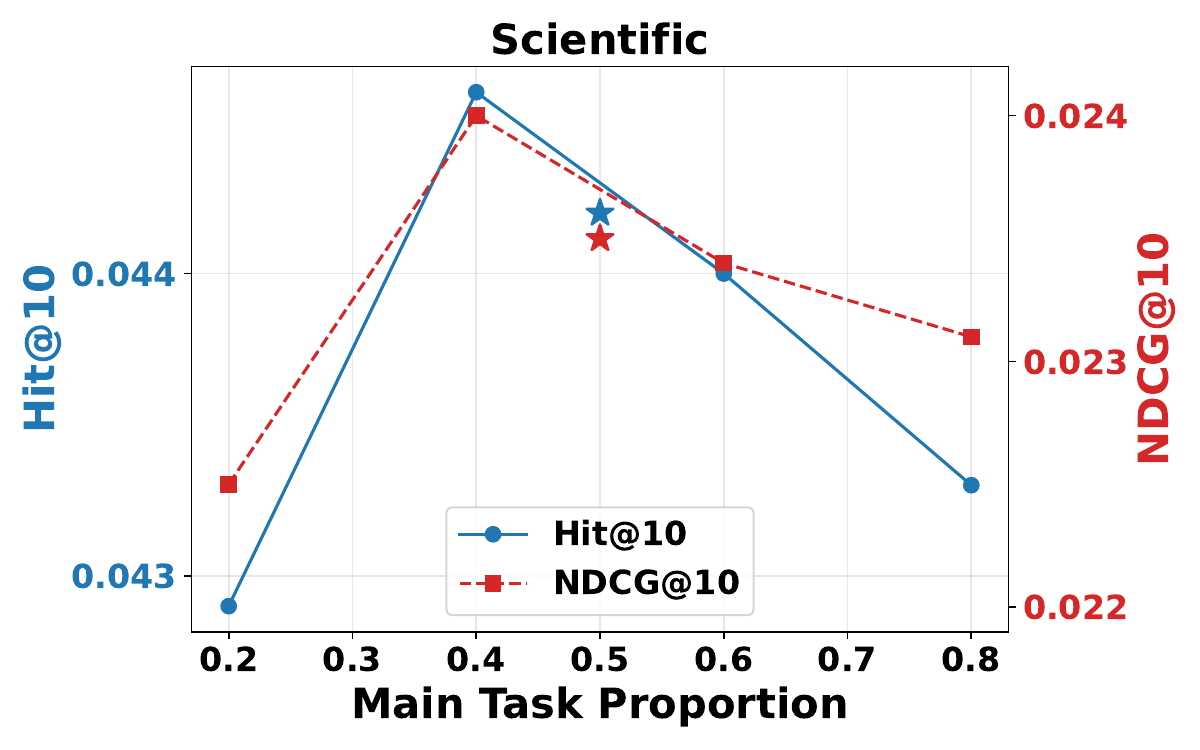}
    \hfill
    \includegraphics[width=0.32\linewidth]{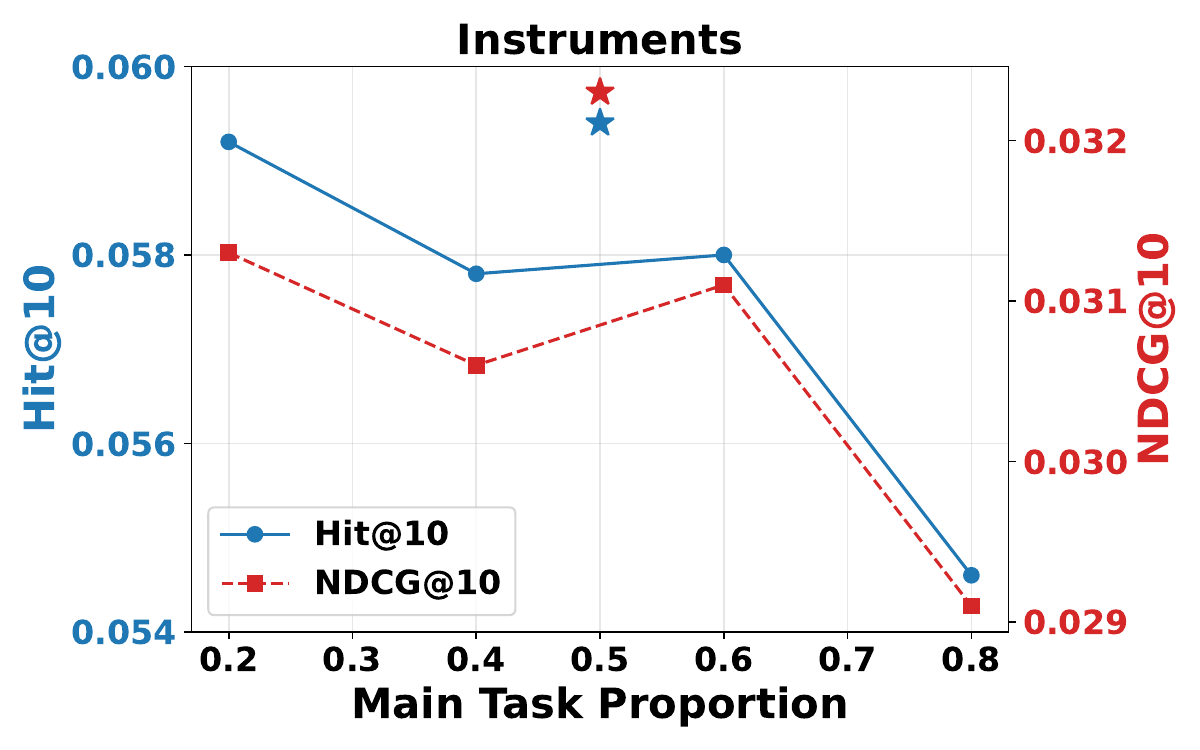}
    \hfill
    \includegraphics[width=0.32\linewidth]{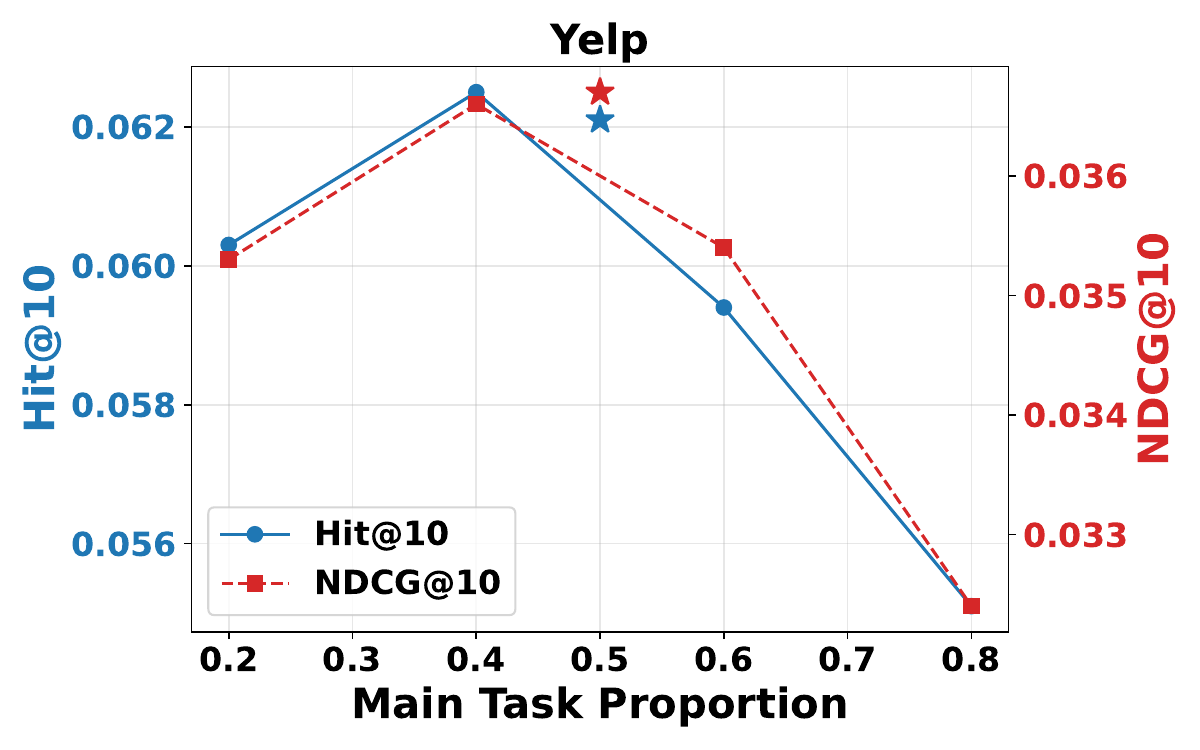} \\
    \caption{The
	    performance of different task proportion
    between the main task and the auxiliary task during training. The number of candidate set $n_c$ is fixed as $10$.
    Our default deterministic setting corresponds to an equal proportion between the two tasks,
    and is indicated by the star markers as reference points (blue as Hit@10 and red as NDCG@10).}
    \label{fig:diff_proportion}
\end{figure*}

In our main experimental setup, each training sequence
participates in both the main and auxiliary tasks within
each training epoch, resulting in an identical proportion for both tasks.
To further investigate the relative importance
of the two tasks,
we consider that
each sample is
assigned to either the main or auxiliary task via
Bernoulli sampling, where the main task is selected with
probability $p$, which corresponds to the main task proportion.
A larger value of $p$ indicates
the model has greater emphasis on the main task.


As shown in Figure~\ref{fig:diff_proportion}, we observe that the performance decreases across the three examined datasets when the main task proportion is disproportionately higher than that of the auxiliary task. Moreover, on the \textit{Scientific} and \textit{Yelp} datasets, the performance also decreases when the main task proportion is substantially lower. These observations indicate that the main task proportion should not be too high, while an excessively low proportion may also be harmful for certain domains. Although the optimal value may vary across different models and datasets, our findings suggest that balancing the two tasks during the learning process remains the most effective approach for achieving good performance on new datasets.

\subsubsection{\textbf{Different Candidate Sampling Strategies} (RQ3)}
\label{sec:exp:ablation:diff-sampling}

In our main setting, we adopt uniform sampling to construct the candidate set.
To better understand how different
sampling strategies affect model behavior, we consider four different candidate
construction settings for the in-depth analysis:
\begin{itemize}[leftmargin=*]
    \item \textbf{Uniform:} Negative candidates are derived by the original
uniform sampling strategy.
    \item \textbf{Long-tail:} Negative candidates consist entirely of long-tail items with the same definition as in Section~\ref{sec:exp:ablation:long-tail}.
    \item \textbf{Warm-set:} Negative candidates consist entirely of warm-set items, defined as non-long-tail items.
    \item \textbf{Dynamic:} Negative candidates are sampled using a dynamic strategy that gradually shifts from long-tail items to warm-set items. Specifically,
    the Bernoulli probability of sampling
    from long-tail items is
    $p = e^{-e_p/T}$,
    with the corresponding probability of sampling from warm-set items given by
    $1-p$, where $e_p$ represents the current
    epoch and
    $T$ is a hyperparameter, set to $100$.
\end{itemize}

\begin{table*}[!t]
\centering
\caption{The overall performance for different candidate
sampling strategies. ($n_c=10$)}
\label{tab:diff-cand-samples}
\vspace{-9pt}
\begin{tabular}{lcccccccc}
\toprule
\multirow{2}{*}{Methods} &
\multicolumn{2}{c}{Scientific} &
\multicolumn{2}{c}{Instruments} &
\multicolumn{2}{c}{Yelp} &
\multicolumn{2}{c}{MovieLens}
\\
\cmidrule(lr){2-9}
& Hit@10 & NDCG@10 & Hit@10 & NDCG@10
& Hit@10 & NDCG@10 & Hit@10 & NDCG@10 \\
\midrule
Uniform
& 0.0442 & 0.0235
& 0.0594 & 0.0323
& 0.0621 & 0.0367
& 0.2770 & 0.1586 \\
Long-tail
& 0.0445 & 0.0239
& 0.0594 & 0.0321
& 0.0629 & 0.0368
& 0.2664 & 0.1517 \\
Warm-set
& 0.0452 & 0.0243
& 0.0588 & 0.0318
& 0.0604 & 0.0353
& 0.2737 & 0.1588 \\
Dynamic
& 0.0458 & 0.0247
& 0.0590 & 0.0320
& 0.0623 & 0.0364
& 0.2740 & 0.1602 \\

\bottomrule
\end{tabular}
\end{table*}
As shown in Table~\ref{tab:diff-cand-samples},
except for the dense \textit{MovieLens} dataset,
we observe that long-tail sampling achieves comparable performance
to uniform or warm-set sampling.
This indicates that incorporating the multiple-choice auxiliary task helps the
model establish better comparative relationships
between sequences and long-tail items, and that the
learned compositions may further generalize to other negative signals.
On the other hand, on the \textit{MovieLens} dataset, warm-set or uniform sampling yields
better performance than long-tail sampling. This could be because the main task
training sequences already contain sufficient long-tail items,
allowing the model to learn the comparative signals between positive items and long-tail negatives. However, the model is more likely to confuse positive items
with warm-set negatives when it is trained only on the main task.
Therefore, separating these warm-set negatives becomes more crucial.

In our experiments, we also examine the dynamic setting
by gradually shifting from long-tail sampling to
warm-set sampling. Under this strategy, we observe that
introducing warm-set negatives in the later stages
can successfully improve the performance over pure long-tail
sampling on the \textit{MovieLens} dataset.
However, across all datasets, the resulting performance does not yield
significant gains compared to the uniform
setting.
This finding suggests that
the benefits from dynamic scheduling are
marginal, and a static uniform sampling strategy
may already be sufficient.

\subsubsection{\textbf{Investigation on the Multiple-Choice Task} (RQ4)}

Since we treat the multiple-choice task as an auxiliary objective and do not include it in the validation phase (Section~\ref{sec:exp:data-and-model:model-imple}),
it is interesting to investigate whether the model
develops the ability to perform the multiple-choice task,
and how it balances information between the interaction history
and the candidate items.
In this section, we further analyze these questions
through the model's selection capability
and its cross-attention behavior in the auxiliary task.
To this end, we construct candidate sets of size $n_c = 2$ 
and $n_c = 10$ for each test instance.

For the selection capability, we
measure whether the top-1 predicted item lies within the candidate set.
As shown in Table~\ref{tab:hit_selection}, we observe that
the prediction accuracy across all datasets for both $n_c = 2$
and $n_c=10$ are equal or close to $1.0$.
This indicates that the model has effectively learned
to conduct the multiple-choice selection when needed.

\begin{table}[!t]
\centering
\caption{Accuracy of whether 
the top-1 predicted item lies within the candidate set.}
\label{tab:hit_selection}
\vspace{-9pt}

\begin{small}
\begin{tabular}{lcccc}
\toprule
Datasets & Scientific & Instruments & Yelp & MovieLens  \\
\midrule
TIGER+$n_c=2$ &  
1.00000 & 1.00000 & 0.99997 & 0.99983 \\
TIGER+$n_c=10$ &
1.00000 & 1.00000 & 0.99984 & 1.00000 \\
\bottomrule
\end{tabular}
\end{small}
\end{table}

\begin{figure}[!t]
    \centering
    \includegraphics[width=0.48\linewidth]{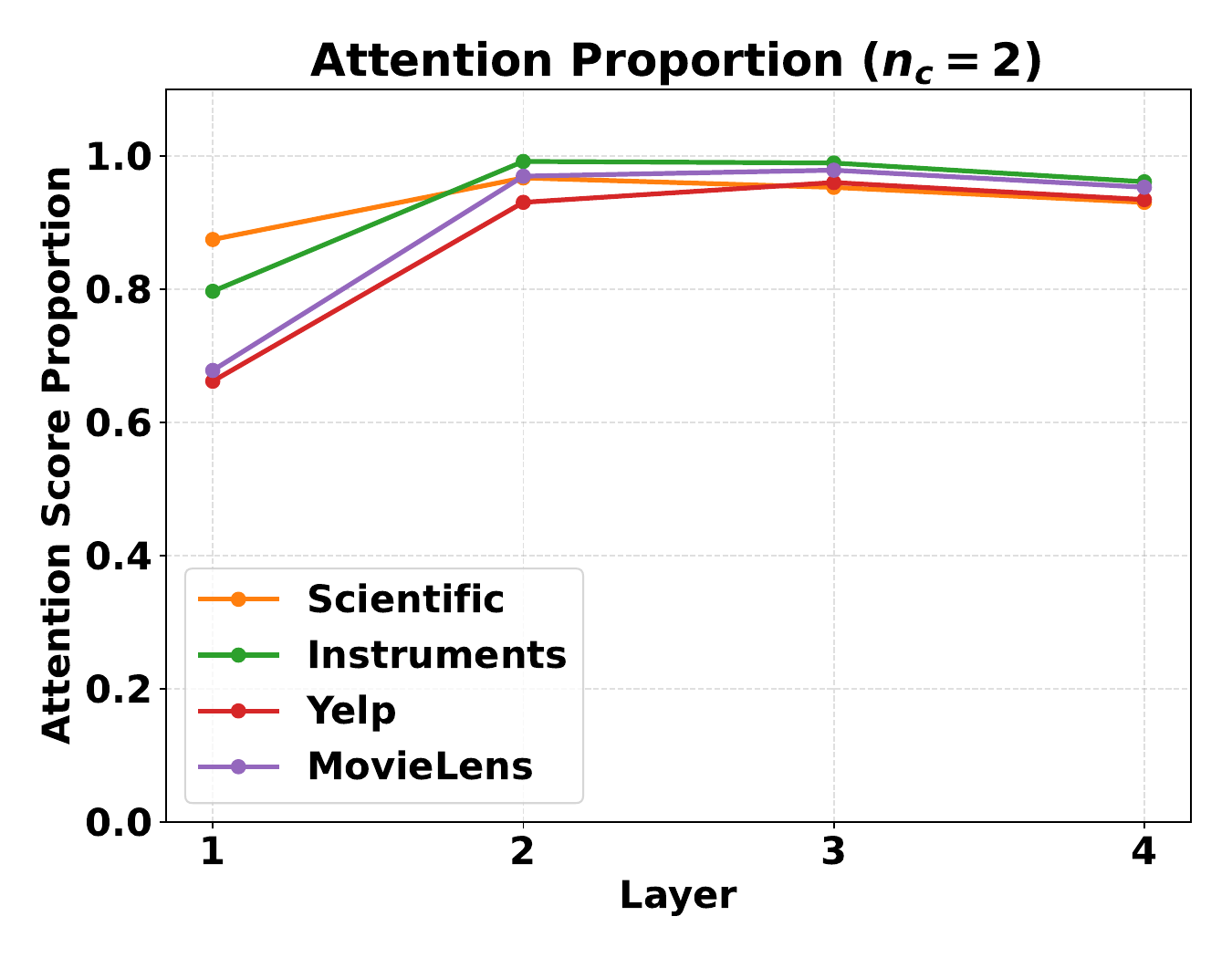}
    \includegraphics[width=0.48\linewidth]{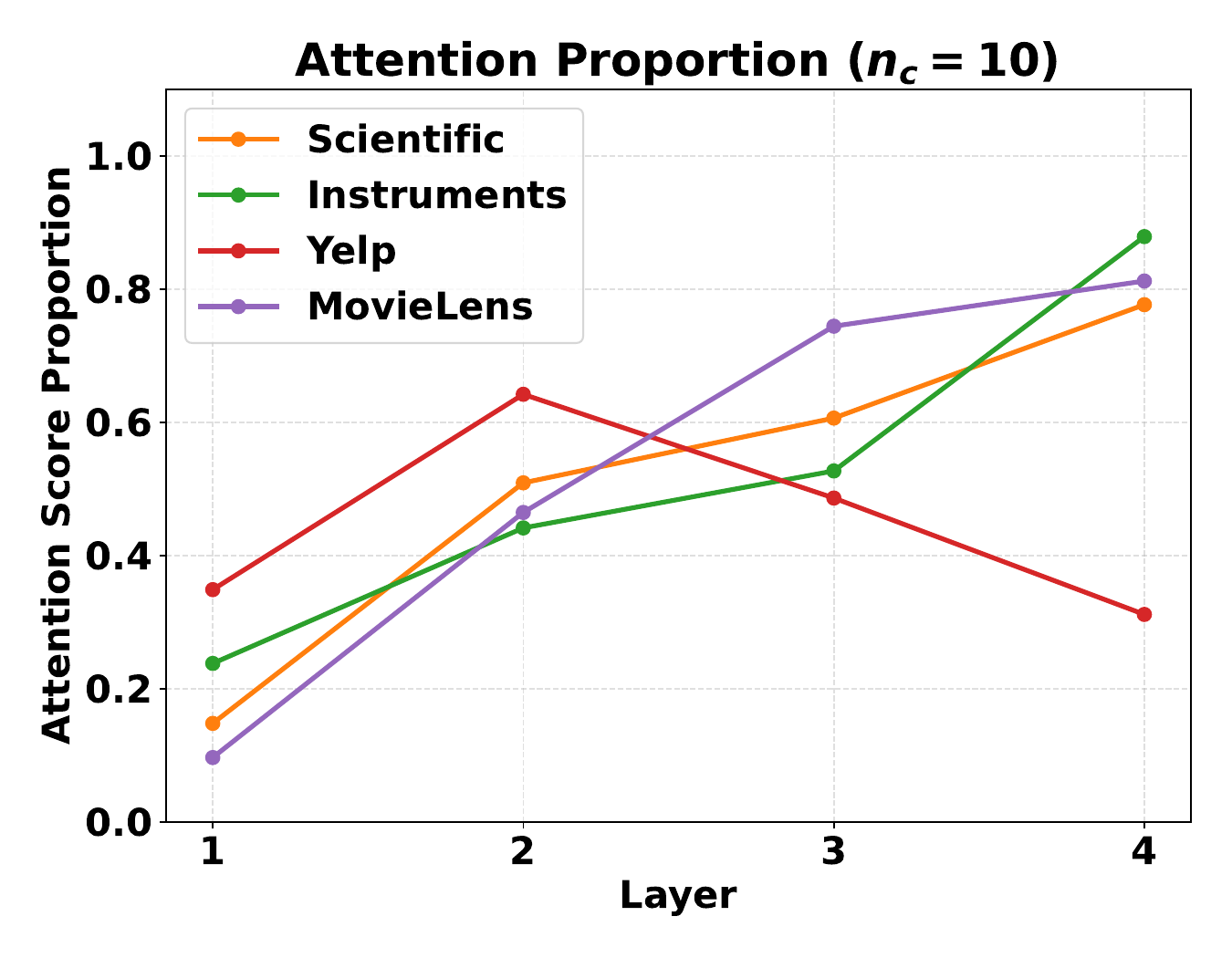}
    \caption{The candidate attention proportion (Equation~\ref{eq:atten_prob}) for $n_c=2$
    and $n_c=10$. The attention score is averaged
    across all heads and tokens in each segment.}
    \label{fig:attention_proportion}
\end{figure}

For the cross-attention 
analysis, we analyze how the decoder starting token
attends to different parts
of the encoder sequence.
We split each encoder sequence into two segments,
the history part (from $p_2$ to $p_3$) and the candidate part
(after $p_3$). We then compute the average 
attention across all heads
and tokens in each segment.
Afterwards, we calculate the
attention proportion of the candidate part
via:
\begin{equation}
\mathrm{cand\_attention\_prop} = \frac{\mathrm{cand\_attention}}{\mathrm{cand\_attention} + \mathrm{history\_attention}} \, .
\label{eq:atten_prob}
\end{equation}
As shown in Figure~\ref{fig:attention_proportion},
we find that in both
$n_c=2$ and $n_c=10$, models
generally
extract more knowledge from the history in 
the early layers, and then become
focused on the candidate knowledge in the
later layers. 
For \textit{Yelp} with $n_c=10$, 
however, the candidate attention 
proportion significantly declines in later layers. 
This potentially stems from candidate information already being integrated into historical representations through encoder self-attention.

\section{Conclusion}
In this paper, we integrate the
multiple-choice task as an auxiliary
into generative recommendation models to alleviate the 
lack of comparative signals. 
Through analysis from the perspectives of 
information bottleneck and compositional 
generalization,
we argue that the multiple-choice 
auxiliary task
can improve the recommendation
model
by encouraging a more robust representation
function
and promoting richer compositional
patterns. 
We conduct extensive experiments 
on four real-world datasets 
and demonstrate that our simple and elegant
learning framework achieves
competitive performance
against
mainstream strong baselines. 

Our work identifies the fundamental 
challenges of generative retrieval with SIDs, 
which provides an alternative direction for improving 
generative recommenders. 
Future work can explore extending the proposed 
framework to different generative architectures, 
as well as developing advanced
domain dependent candidate construction mechanisms.

\appendix
\section*{Appendix}

\begin{table}[!t]
\centering
\caption{The mean and standard deviation of the performance across different dataset for TIGER and TIGER+$n_c=2$. 
We computed 5 runs using different
random seeds. 
The improvements from TIGER+$n_c=2$ to
TIGER are all significant ($p<0.05$)
via Welch's t-test on per-run performance.}
\label{tab:metrics_multiple_seed}
\vspace{-9pt}

\begin{small}
\begin{tabular}{llcc}
\toprule
Model & Datasets & Hit@10 & NDCG@10 \\
\midrule
\multirow{4}{*}{TIGER}
&
Scientific &  
$ 0.0400 \pm 0.0008 $ & $ 0.0213 \pm 0.0004 $  \\
&
Instruments 
& $ 0.0536 \pm 0.0013 $ & $ 0.0284 \pm 0.0006 $ \\
&
Yelp 
&  $ 0.0352 \pm 0.0029 $ & $ 0.0182 \pm 0.0016 $ \\
&
MovieLens 
& $ 0.2630 \pm 0.0050 $ & $ 0.1517 \pm 0.0028 $ \\
\midrule
\multirow{4}{*}{TIGER+$n_c=2$}
&
Scientific &  
$ 0.0442 \pm 0.0012 $ & $ 0.0238 \pm 0.0007 $  \\
&
Instruments 
& $ 0.0581 \pm 0.0006 $ & $ 0.0311 \pm 0.0004 $ \\
&
Yelp 
&  $ 0.0498 \pm 0.0038 $ & $ 0.0282 \pm 0.0024 $ \\
&
MovieLens 
& $ 0.2738 \pm 0.0041 $ & $ 0.1582 \pm 0.0026 $ \\
\bottomrule
\end{tabular}
\end{small}
\end{table}

\begin{table}[!t]
\centering
\caption{The number of
users in different splits 
based on the
frequency of ground-truth items 
(Section~\ref{sec:exp:ablation:long-tail}).}
\label{tab:stats-of-frequency}
\vspace{-9pt}

\begin{small}
\begin{tabular}{lcccc}
\toprule
Dataset
& Scientific & Instruments & Yelp & MovieLens  \\
\midrule
popular &
29835 & 39632 & 18336 & 4518 \\
medium &
9456 & 7156 & 5287 & 1198 \\
long-tail &
11694 & 10651 & 6808 & 324 \\
\bottomrule
\end{tabular}
\end{small}
\end{table}

\section{More Details on Experiments}
\label{sec:append:more-exp}

We further report the mean
and standard deviation 
of the overall performance
for TIGER and ``TIGER+$n_c=2$'' across 
5 different random seeds in
Table~\ref{tab:metrics_multiple_seed},
as well as the testing
subset statistics for the
analysis of predicting items
with different frequency (Section~\ref{sec:exp:ablation:long-tail}) in Table~\ref{tab:stats-of-frequency}.

\bibliographystyle{ACM-Reference-Format}
\bibliography{references.bib}

\end{document}